# Title: Early Mars' habitability and global cooling by $H_2$-based methanogens

**Author list:**

**Boris Sauterey[1,2]***, Benjamin Charnay[3], Antonin Affholder[2,4], Stéphane Mazevet[5,§], Régis Ferrière[1,2,6,§]

**Affiliations:**

[1]Department of Ecology & Evolutionary Biology, University of Arizona; Tucson, AZ 85721, USA.

[2]Institut de Biologie de l'Ecole Normale Supérieure (IBENS), Université Paris Sciences et Lettres, CNRS, INSERM; 75005 Paris, France.

[3]LESIA, Observatoire de Paris, Université PSL, CNRS, Sorbonne Université, Université de Paris; 5 place Jules Janssen, 92195 Meudon, France.

[4]Insitut de Mécanique Céleste et de Calcul des Éphémérides, Observatoire de Paris - PSL; 75014 Paris, France

[5]Observatoire de la Côte d'Azur, Université de la côte d'Azur, Boulevard de l'Observatoire, 06304 Nice, France.

[6]iGLOBES International Research Laboratory, CNRS, Ecole Normale Supérieure, Université Paris Sciences et Lettres, University of Arizona; Tucson, AZ 85721, USA.

[§]Co-senior authors

*Corresponding author. Email: boris.sauterey@biologie.ens.fr

**Introductory paragraph:**

**During the Noachian, Mars' crust may have provided a favorable environment for microbial life[1,2]. The porous brine-saturated regolith[3-5] would have created a physical space sheltered from UV and cosmic radiations and provided a solvent, while the below-ground temperature[2] and diffusion[6,7] of a dense reduced atmosphere[8,9] may have supported simple microbial organisms that consume $H_2$ and $CO_2$ as energy and carbon sources and produce methane as a waste. On Earth, hydrogenotrophic methanogenesis was among the earliest metabolisms[10,11] but its viability on early Mars has never been quantitatively evaluated. Here we present a probabilistic assessment of Mars' Noachian habitability to $H_2$-based methanogens, and quantify their biological feedback on Mars' atmosphere and climate. We find that subsurface habitability was very likely, and limited mainly by the extent of surface ice coverage. Biomass productivity could have been as high as in early Earth's ocean. However, the predicted atmospheric composition shift caused by methanogenesis would have triggered a global cooling event, ending potential early warm conditions, compromising surface habitability and forcing the biosphere deep into the Martian crust.**

**Spatial projections of our predictions point to lowland sites at low-to-medium latitudes as good candidates to uncover traces of this early life at or near the surface.**

**Main Text:**

To assess and quantify the habitability of early Mars and the evolution of its surface conditions under the influence of methanogenic hydrogenotrophy, we use a state-of-the-art 1D photochemical-climate model combined with a crust model to self-consistently compute the atmospheric chemical composition, climate, thermal profile of the crust, and crust-atmosphere gas exchanges (Methods, Fig. 1, Extended Data Fig. 1). The geophysical-chemical model is coupled with a depth-structured model of chemoautotrophic ecosystem adapted from our previous work[12,13] to (i) evaluate the habitability of the Martian subsurface to populations of methanogenic hydrogenotrophs, (ii) resolve the dynamics of these populations, and (iii) quantify the corresponding biological feedback on the planet's atmosphere and climate.

Only the fraction of the surface left free of ice (measured by $\boldsymbol{\rho}$) allows crust-atmosphere gas exchanges, as a consequence of ice forming in the regolith pores blocking gas pathways into the crust[6]. We assume that the planet's ice coverage is determined by surface temperature and the freezing point of saline brines (Methods, Extended Data Fig. 2). The brines freezing point is poorly constrained (estimates range from 203 to 273 K) as it depends on the brines composition[3-5]. We therefore simulated initial and steady state characteristics of Mars for three values of freezing point –203, 252 and 273 K– corresponding respectively to perchlorate brines, NaCl brines, and pure water[4].

From our climate-atmosphere model (Fig. 1) we resolve Mars' initial features (Fig. 2A-C). The initial distribution of average surface temperature, $\overline{T}_{surface}$, ranges from 216 to 294 K, with a median at 256 K (Fig. 2C). Below the surface, a general pattern is that temperature increases with depth while diffusivity decreases. For brines freezing at 203, 252 and 273 K, Mars is fully covered with ice (hence uninhabitable) in 0, 10 and 40% of the cases, with median $\boldsymbol{\rho}$ of 100%, 75%, and 0.15% respectively (Fig. 2D). Thus, the nature of the brines would have strongly constrained the geographic extent of Mars initial habitability. Note that our model shares with other recent models of Mars climate[15,16] the assumption that albedo remains low and constant in spite of changing ice coverage (see Methods). We will further discuss this assumption and argue that our results are conservative with respect to the triple interaction between surface temperature, albedo, and biological activity.

The atmospheric redox disequilibrium and its accessibility in the Martian crust make the hydrogenotrophic ecosystem viable in all of our simulations in which Mars is not fully covered with ice. Consistently with microbial data from permafrost ecosystems on Earth[17] our model predicts that the lowest temperature for hydrogenotrophs to exist and reproduce is ~253 K (Fig. 2E-G). For warmer surface temperature, methanogens can colonize the first layer of the Martian crust. When the surface is colder, methanogens are limited upward at the depth at which the limit temperature of 253 K is reached (see Methods). The deeper end of the microbial vertical distribution is bounded *ca.* 320 K (lower than the maximum temperature at which methanogenic extremophiles can grow on Earth[18]). At such depths the atmospheric redox potential diffusing from the surface has been entirely exploited by the ecosystem l above.

As they colonize the Martian subsurface, methanogenic hydrogenotrophs drive atmospheric $CH_4$ up and atmospheric $H_2$ down. At steady state, the biogenic rates of $CH_4$ production and $H_2$ consumption, combined with $H_2$ atmospheric escape, balance out the loss rate of $CH_4$ by photochemistry and the production rate of $H_2$ by photochemistry and volcanic outgassing (Fig. 1). Typically, the planetary system reaches this new steady state in 100,000 to 500,000 years. As a result, the median $f_{H_2}$ drops from 5% to between 0.35% and 2.75% depending on the brines freezing point (Fig. 2A) while the median steady state atmospheric concentration in $CH_4$ rises to 0.075% to ~1% (Fig. 2B). Due to the respective effect of $H_2$ and $CH_4$ on climate[14-16] (Fig. 1), the global atmospheric shift triggered by the microbial biosphere drives a strong global cooling effect (Fig. 2C). Even though the biological impact on Mars surface conditions markedly depends on the brines freezing point and initial ice coverage (Fig 2A-D), methanogenic hydrogenotrophs make a warm early Mars unlikely, as the maximum temperature plummets under their planetary influence from 294 K to between 260 and 250 K, across the range of brines freezing point values (Fig. 2C).

Reciprocally, the global cooling feeds back to the biosphere. First, when the average surface temperature in ice-free regions drops below 253 K, methanogens are forced deeper into the Martian crust (Fig. 2E-G). Second, as the biosphere uses up the atmospheric redox potential, the atmospheric change entails that the planet's global thermodynamic favorability to methanogenesis declines. As a consequence, the total biomass productivity at steady state falls hundred-fold (Fig. 2H). Finally, the ice-free (hence potentially habitable) regions of Mars can shrink dramatically (Fig. 3 and Supplementary Video 1). For example, for brines freezing at 252 K the median $\boldsymbol{\rho}$ drops from 83% to 2% under the influence of methanogens (Fig. 2D). In spite of this biologically induced reduction of Mars habitability, the predicted planetary biomass production at steady state is similar to biomass production estimates for the same ($H_2$-based methanogenic) primitive biosphere in the Archean Earth's ocean[12,19].

The best validation of our predictions would come from the discovery on present-day Mars of methanogenic life descending from the early metabolism modeled here. When run for atmospheric conditions corresponding to modern Mars our model predicts the atmosphere to be an insufficient source of electron donors for $H_2$-based methanogens to survive. As Mars' atmosphere became thinner during the Hesperian and early Amazonian, a putative biosphere persisting throughout the Noachian would have had to shift its main energy source from the vanishing atmospheric redox gradients to hydrothermal or radiolytic ones, deeper in the Martian crust. Deep chemotrophic ecosystems exist on Earth; an extant ecosystem on modern Mars might be of that kind[22,23], and could explain the repeated[23] yet highly debated[24] detection of $CH_4$ traces in the lower Martian atmosphere. Our model could be adapted to quantify the habitability of modern Mars' crust to such ecosystems and constrain their depth and productivity.

In the meantime, our model can help inform the search for fossilized biomarkers of Noachian $H_2$-based methanogens. Among the many types of biosignatures that have been proposed to identify ancient metabolic activity[2,20,25,26], isotopic fractionation seems to be the most reliable and commonly used[26]. Detecting isotopic signatures of anaerobic chemotrophic life on Noachian Mars and even of specific metabolic activity (e.g., methanogenesis) might indeed be possible in the light of the analysis of 3-to-4 Gy-old Archean fossils from terrestrial Mars-analogs (reviewed in ref[2, 26]). Near-surface populations would have been the most productive ones (Fig. 2 E-G), therefore maximizing the likelihood of biomarkers preserved in detectable quantities. The first few meters of the Martian crust are also the most easily accessible to exploration given the technology currently embarked on Martian rovers. The probability of life traces at or near the surface is strongly dependent on the freezing point of Martian brines (Fig. 2E-G); to identify regions where this probability is highest, and evaluate current exploration sites such as Isidis Planitia and Jezero crater, we performed spatial projections of the model outputs (Fig. 3 and 4, Extended Data Fig. 3; see Methods).

Temperature drops with latitude and elevation[27] (see Methods). Thus, initial ice coverage was most likely in poleward regions and highlands and least likely in lowlands located at low-to-medium latitudes such as Isidis and Hellas Planitiae and in the Noachian lakes scattered along the North-South dichotomy[28] (including Jezero crater). With a low brines freezing point (203 K), the Martian surface is initially ice-free and globally habitable. As the early biosphere expands, the biological feedback is massive and cools the planet down dramatically (Fig. 2C), leaving the surface ice-free (due to the low brines freezing point) but driving the biosphere deep in the crust (Fig. 2E). The probability of near-surface methanogenic life at steady-state is ca. 0.5 in Hellas Planitia, 0.2 in Isidis Planitia and less than 0.15 at Jezero Crater (Fig. 4A; see Extended Data Fig. 3 for median minimum depth of methanogenic life). With an intermediate or high brines freezing point (252 K or 273 K), a significant fraction of the Martian surface may have been initially frozen (Fig. 3), reducing the habitable region but also the biological feedback to climate. At life-atmosphere steady state, the relatively high brines freezing point and relatively weak biological feedback balance out with a large ice coverage yet relatively warm ice-free surface, resulting in methanogenic life being limited to small ice-free regions but at or very close to the surface. With brines freezing at 252 K, Hellas and Isidis Planitiae and Jezero Crater are among the few regions that may remain free of ice, where the median minimum depth of Noachian methanogens would then be within the top few meters (Extended Data Fig. 3). The probability of near-surface life reaches 1 in some areas of Hellas Planitia, 0.5 in Isidis Planitia and 0.3 at Jezero Crater (Fig. 4B). With brines freezing at 273 K, the biological feedback, albeit weak, would have caused Isidis Planitia and Jezero Crater to freeze, thus compromising these areas' habitability. Only Hellas Planitia would have remained likely ice-free at steady state, in which case the probability of near-surface habitability would have been of one (Fig. 4C and Extended Data Fig. 3).

We conclude that, somewhat counterintuitively, the condition that makes Mars initial habitability to methanogens least likely (high freezing point of Martian brines), is also the condition under which signs of early Martian methanogenesis might be easiest to detect today. With early Mars' brines freezing point high enough, Hellas and Isidis Planitiae and Jezero Crater appear to encompass the best candidate sites to search for signs of methanogenic life that might have persisted near the surface throughout the Noachian. Access to these biomarkers may, however, be obstructed by the accumulation of sediments from the Late Hesperian and later[29,30], unlikely to have trapped later biomarkers because life would have, by then, either disappeared or migrated deep into the crust.

Habitability and climate feedback of hydrogenotrophic methanogens were recently quantified for the early Archean Earth[12]. The results reported here for the Noachian Mars show striking similarities and differences. On the one hand, models predict very likely habitability to hydrogenotrophic methanogens on both young planets, with similar biomass production. On the other hand, climate feedbacks work in opposite directions. While hydrogenotrophic methanogens may have contributed to maintaining temperate conditions on Earth[12,19], they would have cooled the early Martian surface, with a reduction of the maximum possible temperature by 33 to 45 K. Such divergence in climate evolution is the consequence of different prebiotic atmospheric compositions. For a $CO_2$-dominated atmosphere as on early Mars, $H_2$ has a stronger greenhouse effect (from $CO_2$-$H_2$ collision-induced absorptions) than $CH_4$, in contrast to a $N_2$-dominated atmosphere as on the early Earth[14-16]. Contrasted planetary responses of Earth and Mars to metabolic activity might have occurred repeatedly as their biospheres would have evolved and diversified. On Earth, the evolution of methanotrophy (biological consumption of methane) could have transiently offset the warming effect of methanogenesis[12]; in contrast, methanotrophy could have driven a warming event on the Late Noachian/Hesperian Mars. The co-evolution of Martian surface conditions with a diversifying methane cycling biosphere therefore adds to mechanisms possibly explaining an early global cooling event as well as transient warming episodes that seem to have punctuated the early Martian climatic history[8].

Our model does not take into account the biology-to-climate feedback specifically driven by albedo increasing with surface ice formation. As a consequence, our model likely underestimates the cooling effect of hydrogenotrophic methanogens on early Mars[14]. This additional feedback could have amplified the direct atmospheric impact of methanogenesis on Mars climate and triggered a global glaciation. Although quantification of this effect warrants further development of Mars climate models[14-16], the mechanism in and of itself points to the possibility that life-environment feedbacks can compromise habitability at planetary scale. This Medean scenario[31] (self-destructive life-planet feedbacks) adds conceptually to Gaian bottlenecks (life-planet feedback failing to counter the geophysical loss of habitability[32]) as a potential limit to the long-term habitability of planets and planetary bodies in our solar system and beyond.

## Methods

Atmospheric and climate model

To evaluate the photochemical destruction/production of $H_2$ and $CH_4$ (the two main atmospheric species in our planetary ecosystem model) we ran the latest version of the 1D photochemical model of Mars developed in the Virtual Planetary Laboratory[33], on grids of atmospheric compositions. The results are in line with previous estimates[34], and based on them, we interpolated the photochemical rates as functions of the atmospheric composition (Fig. 1A and B).

Similarly, we evaluated the dependence of the Martian climate on its atmospheric composition and pressure based on the Virtual Planetary Laboratory's 1D climate model, updated to account for the latest data on the respective collision-induced warming effect of $H_2$ and $CH_4$[14-16]. Again, we used the outputs of the model to interpolate the mean surface temperature as a function of the atmospheric composition ($f_{H_2}$, $f_{CH_4}$, $f_{CO_2}$) and total atmospheric pressure (Fig. 1C-E). Because of the collision induced absorption resulting from the $CO_2$-saturated atmosphere of early Mars, $H_2$ appears to be a more potent greenhouse gas than $CH_4$, in line with previous studies[14-16]. Methane may even produce an anti-greenhouse effect at low atmospheric pressure and low $f_{H_2}$.

Following previous work[15,16] the climate model was run assuming a constant low planetary albedo of 0.2 corresponding to low ice coverage. Note however that a low temperature is expected to be associated with an extended ice coverage (see the Methods, "spatial projections" section) and a high planetary albedo. It was previously shown[14] that a high planetary albedo can cool Mars climate down considerably. Therefore, in the case of Mars global cooling, the resulting expansion of surface ice and increase in planetary albedo may have mediated a positive feedback strengthening the cooling event and potentially triggering a cold runaway scenario. To our knowledge, this climate-surface ice feedback loop remains to be integrated and evaluated dynamically in climate models of Mars.

Crust model

Subsurface temperature is expected to increase with depth, while the availability of diffusing atmospheric gases drops. Following on ref[7], we assume a linear temperature gradient with depth, starting from the surface temperature and warming as depth increases:

$$T(z) = T_{surface} + a_T z \tag{E1}$$

with $z$ the depth in kilometers and $a_T$ the rate at which temperature increases with depth ranging between 10 and 40 K km$^{-1}$ [7]. The decrease with depth of the diffusivity (in cm$^2$ s$^{-1}$) in the water-saturated crust depends partially on the temperature depth-profile, according to ref[7]:

$$D_i(z) = \frac{\epsilon(z)\, r(z)}{3\, \tau(z)} \sqrt{\frac{8\, RT(z)}{\pi m_i}} \tag{E2}$$

with $i$ the considered chemical species $X_i$ ($X_i$ = $H_2$, $CO_2$, $CH_4$, or $N_2$), $m_i$ its molar mass, $\epsilon(z)$ the porosity of the crust, $r(z)$ the pores' radius (in cm), and $\tau(z)$ the tortuosity. The pore radius $r(z)$ follows a linear decrease with depth $r(z) = r(0) - a_r z$ with $a_r = \frac{r(0)}{z_{max}}$ so that $r(z) = 0$ when $z = z_{max}$, the depth of pore closure. The crust porosity and tortuosity both follow an exponential decrease with depth, respectively $\epsilon(z) = \epsilon(0)e^{-\frac{z}{z_{max}}}$ and $\tau(z) = \tau(0)e^{-\frac{z}{3z_{max}}}$. The vertical flux (in molecules cm$^{-2}$ s$^{-1}$) can then be written as

$$F_i = D_i(z)\frac{\partial n(X_i)}{\partial z} \tag{E3}$$

$n(X_i)$ being the density of species $X_i$ (in molecules cm$^{-3}$). The envelopes and distribution of depth profiles for the temperature and diffusivity of $H_2$ corresponding to the soil parameter ranges explored (see Extended Data Table 1) are shown in Extended Data Fig. 1.

Ecological model

Our ecological model describes the dynamics of biological populations of chemotrophic cellular organisms. Equations for the growth and death of individual cells are derived from how, in each individual cell, energy flows from catabolism (energy acquisition) to anabolism (cell maintenance first, then biomass production). The individual metabolism is described by

$$\text{Catabolism:} \quad \sum_{i=1}^{n} \gamma_{S_i}^{cat} S_i \xrightarrow{\Delta G_{cat}} \sum_{i=1}^{m} \gamma_{P_i}^{cat} P_i$$

$$\text{Biomass production:} \quad \sum_{i=1}^{n'} \gamma_{S_i}^{bio} S_i \xrightarrow{E_{bio}} \sum_{i=1}^{m'} \gamma_{P_i}^{bio} P_i \qquad \text{(E4)}$$

where $S_i$ and $P_i$ are the substrates and products of the metabolic reactions that are specific to the considered metabolism and the $\gamma$'s their stoichiometric coefficients, and $\Delta G_{cat}$ and $E_{bio}$ the energy released by the catabolic reaction and necessary to biomass production, respectively. In the case of hydrogenotrophic methanogens, the catabolic reaction is $CO_2 + 4\ H_2 \rightarrow CH_4 + 2\ H_2O$. The value of $\Delta G_{cat}$ is given by the Nernst relationship

$$\Delta G(T) = \Delta G_0(T) + RT\log(Q) \qquad \text{(E5)}$$

where $R$ stands for the ideal gas constant, $T$ for temperature (in K), $\Delta G_0(T)$ (in kJ mol eD$^{-1}$) for the standard Gibbs free energy of the reaction, and $Q$ for the reaction quotient $\frac{\prod_{i=1}^{n} S_i^{\gamma_{S_i}}}{\prod_{i=1}^{m} P_i^{\gamma_{P_i}}}$. The value of $\Delta G_0$ is obtained from the Gibbs-Helmholtz relationship:

$$\Delta G_0(T) = \Delta G_0(T_S)\frac{T}{T_S} + \Delta H_0(T_S)\frac{T_S - T}{T_S} \qquad \text{(E6)}$$

where $T$ is the temperature of the medium, $T_S$ the standard temperature of 298.15 K, and $\Delta H_0(T_S)$ the standard enthalpy. The catabolic acquisition of energy occurs at a rate $q_{cat}$ (in mol eD cell$^{-1}$ d$^{-1}$). The energy obtained is first directed toward maintenance, with $E_m$ (in kJ cell$^{-1}$ d$^{-1}$) the biomass specific energy requirements for maintenance per unit of time. The energy requirements of the cell can be expressed in terms of the rate at which the catabolic reaction must occur for the cell to function, $q_m$ (in mol eD cell$^{-1}$ d$^{-1}$), with:

$$q_m = \frac{-E_m}{\Delta G_{cat}}. \tag{E7}$$

The cell maintenance requirements are met when $q_{cat} > q_m$. If they are not (i.e., $q_{cat} < q_m$), a decay related term $k(q_m - q_{cat})$ (in d$^{-1}$) is added to the basal cellular mortality rate, $m$ (in d$^{-1}$), with $d = k(q_m - q_{cat}) + m$ the effective mortality rate (in d$^{-1}$). If $q_{cat} > q_m$, the energy remaining after maintenance $(q_{cat} - q_m)\Delta G_{cat}$ (kJ cell$^{-1}$ d$^{-1}$) can be allocated to biomass production. The assimilation of each mol of carbon into biomass requires a quantity of energy $E_{bio}$ (in kJ mol $C_{org}^{-1}$) corresponding to the sum of the costs of producing the biomass, $\Delta G_{ana}$, and organizing it, $E_{diss}$. The term $E_{diss}$ is a phenomenological estimate and its value is the same as in ref[12] (see Supplementary Table 2). The value of $\Delta G_{ana}$ is obtained from the Nernst equation (equation E5) and assuming the following anabolic reaction: 24 $H_2$ + 10 $CO_2$ + 1 $N_2$ $\rightarrow$ $C_{10}H_{18}O_5N_2$ + 15 $H_2O$. Note that we assume that Martian methanogens would have been able of fixing N from the atmospheric $N_2$, as their terrestrial counterparts most likely were[35]. As an alternative to biological $N_2$-fixation, the atmospheric production and rain-out of bioavailable forms of N (ammonia, nitrate) could have been coupled to the lightning-induced breakdown of atmospheric $N_2$[21,36]. Infiltration of those fixed nitrogen compounds into the subsurface would have been possible during the Noachian, in location when and where there was surface liquid water. Should have biological nitrogen fixation not evolved on Mars, the exclusive use of fixed nitrogen compounds by methanogens would have set additional constrains on underground habitability to methanogenic life. The efficacy of the metabolic coupling $\lambda$ (i.e., the necessary number of occurrences of the catabolic reaction to fuel one occurrence of the reaction of biomass production, in mol eD mol $C_{org}^{-1}$) is the ratio between the energy produced by the catabolic reaction and the energetic cost of biomass production

$$\lambda = \frac{-\Delta G_{cat}}{E_{bio}}. \tag{E8}$$

Biomass is produced at a rate $q_{bio}$ (in mol $C_{org}$ $cell^{-1}$ $d^{-1}$) with

$$q_{bio} = \lambda(q_{cat} - q_m). \tag{E9}$$

The metabolic rates $q_{cat}$ and $q_{bio}$ depend on the concentration of the metabolic substrates through a Michaelis-Menten term:

$$q = q_{max} \frac{S_{lim}}{S_{lim} + K} \tag{E10}$$

where $S_{lim}$ is the most limiting nutrient, $q_{max}$ the maximum reaction rate, and $K$ the half-saturation constant (we assume the same maximum rates and and half-saturation constants for all the reactions, their values being based on phenomenological estimates; see Supplementary Table 2). Rates increase with both cell size $S_C$ (radius in µm) and temperature, which we describe with power laws. Finally, the optimal cell size, i.e., the cell-size that maximizes the individual ability to exploit its environment, is itself estimated through a power law (see ref[12] for details). The default parameter values based on empirical estimates and used in the main text are given in Supplementary Table 2 (details about these values can be found in ref[12]).

A key feature of the model is that it estimates both the thermodynamic and kinetic temperature dependencies of the metabolism. Typically, high temperatures result in a decreased thermodynamic coupling between energy acquisition and biomass production – i.e., an energetic constraint. Lower temperatures on the other hand result in lower metabolic rates – i.e., a kinetic constraint. The metabolic efficiency of $H_2$-based methanogens is therefore optimized at intermediate temperatures, the exact value being set by the ecological context (e.g., fluxes of metabolic substrates and wastes, basal mortality[13]). Note that for the sake of simplicity, we neglect the potential effect of low water activity and osmotic constraints resulting from the high salinity of the Martian hydrosphere on the physiology of methanogens. The effect of these unaccounted constraints on Martian life are hard to evaluate as they are poorly understood on Earth; any assumption on what cellular life's adaptation to such constraints might have been on Mars would be highly speculative. Our model however could be complexified to include the thermodynamic component of these constraints.

Physiological rates can then be implemented in a model of ecological dynamics describing the variation through time of the methanogenic population abundance $B$ (in cells) as the balance of biomass production and death, and the abundance of each chemical species in the medium $X_i$ as the balance of the inward/outward flux $F(X_i)$ between the local ecosystem (here a given point in the crust column) and the exterior (here provided by the crust model; see equation E3) and biological consumption/production:

$$\frac{dB}{dt} = \left(\frac{q_{bio}}{Q_{C_{org}}} - d\right)B$$
$$\frac{dX_i}{dt} = F(X_i) + \left(q_{cat}\gamma_{X_i}^{cat} + q_{bio}\gamma_{X_i}^{bio}\right)B \qquad \text{(E11)}$$

where $Q_{C_{org}}$ is the structural carbon content of a cell. From equation (E11) we obtain a quantitative criterion for habitability, which is that the initial environmental conditions must be compatible with biological population growth, i.e., $\frac{q_{bio}}{Q_{C_{org}}} > d$. By solving for the equilibrium of (E11) we also quantify (i) the biological feedback on the local chemical composition, and (ii) the ability of the local ecosystem to influence the larger scale of the environment through the value of the interaction term $F(X_i)$ at steady state (see equation E3). While the ecological model is similar to the model used in Sauterey et al. (2020)[12], its integration along the spatially structured environmental gradient provided by the crust model is unique to this study. This coupled model provides us with average surface fluxes at the crust-atmosphere interface corresponding to the planetary conditions set by the atmospheric and climate model. These fluxes are then integrated in the global planetary model over the ice-free surface of Mars; they feed back dynamically to the atmosphere and climate, driving the evolution of Mars' surface conditions over time.

Note that we made the assumption, common for chemotrophic ecosystems[12,13,19], that biomass production would have been limited either by the energetic yield of methanogenesis (controlled by the redox imbalance of the atmosphere) or by the availability of the key elements C, H, N and O, obtained from $CO_2$, $H_2$, $N_2$. Although phosphorus and sulfur-containing minerals are abundant on Mars[20,21], whether P and S may have been biologically limiting is currently unknown. Our estimates of carbon assimilation (Fig. 2H) provide a basis to actually quantify the macronutrients abundance that would be needed to sustain the levels of biomass production that the model predicts.

Minimum depth of the hydrogenotrophic methanogenic ecosystems

Due to the kinetic constraints on the $H_2$-based methanogenic metabolism, the lowest viable temperature is approximately 253 K. The minimum depth at which the ecosystem can exist in the Martian crust, $z_{bio}$, can be found by considering the temperature gradient of the crust:

$$z_{Bio} = \max\left(\frac{253 - T_{surface}}{a_T}, 0\right) \tag{E12}$$

where $T_{surface}$ is the surface temperature and $a_t$ the temperature gradient (in K km$^{-1}$).

Probabilistic simulations

We run the coupled crust-ecosystem-atmosphere-climate model in a probabilistic framework, by performing a Monte-Carlo exploration of likely ranges for each planetary parameter in the model (Supplementary Table 1). From the simulations of 3,000 plausible young Mars, we obtain probabilistic estimates of global properties (atmosphere composition, climate, ice coverage, and the productivity and depth profile of the methanogenic biosphere) prior to biologically induced changes to the surface conditions, and after these changes at steady state.

Spatial projections

To obtain spatial projections, we begin with the posterior distributions of average surface temperature $\bar{T}_{surface}$ produced by our model. Then we evaluate the probability distribution of surface temperature at any location given latitude and elevation based on Fastook and Head's model[27]. The authors used simulations from the LMD Generic Climate Model to derive empirical relationships between local surface temperature ($T_{surface}$) and a base temperature, latitude ($lat$), and elevation ($Z$), for various scenarios of atmospheric pressure (0.008, 0.2, and 1 bar). We modified the relationship found for an atmospheric pressure of 1 bar so that the local surface temperature is expressed as a function of the average surface temperature instead of a basal temperature, and obtain

$$T_{surface} = \bar{T}_{surface} - 5\pi + 20\cos\left(\frac{lat\ \pi}{180}\right) + 2.4Z \quad \text{(E13)}$$

Based on a topographic map of Mars[37], this relationship is used to obtain a map of the local temperatures as a function of Mars average surface temperature. We consider that ice covers all the locations at which the local temperature is inferior to the brines freezing point. From that, we draw the maps of the ice coverage corresponding to each of the three values of brines freezing point (Fig. 3 and Supplementary Video 1). We then evaluate the habitable (i.e., ice-free) fraction $\boldsymbol{\rho}$ of Mars as a function of the global average surface temperature for each of the values of brines freezing point (Extended data Fig. 2A). Finally, for each value of the brines freezing point, we integrate the average surface temperature over the habitable fraction of the Mars surface (Extended Data Fig. 2B). The obtained relationships between the global average surface temperature, the habitable fraction of Mars $\boldsymbol{\rho}$, and the average surface temperature in this habitable fraction are then implemented in the global ecosystem model.

Equation (E13) is also used to compute, based on the average steady-state surface temperature distribution obtained from our simulations, temperature distributions at any location on Mars for each of the three scenarios explored. From these distributions and equation (E12) we infer the median minimum depth of the biosphere (i.e., the median depth at which the temperature reaches 253 K, the limit temperature to viability; Fig. 4) and the probability of the biomass reaching the surface (i.e., the probability that the local surface temperature is higher than the 253 K threshold; Extended Data Fig. 3).

**Data availability:**
The datasets produced and analyzed in this study are available in the following repository: https://github.com/bsauterey/MarsEcosys[38] (doi: 10.5281/zenodo.6963348)

**Code availability:**
The planetary ecosystem model coupling climate, atmosphere, ice coverage and belowground ecosystem and the datasets produced with it are available in the following repository: https://github.com/bsauterey/MarsEcosys (doi: 10.5281/zenodo.6963348). The photochemical and climate models are accessible on the Virtual Planet Laboratory's gitlab (https://github.com/VirtualPlanetaryLaboratory/atmos[38]), the adapted versions used in this study are available upon request.

**Acknowledgements:**

We are grateful for discussion with D. Apai, A. Bixel, Z. Grochau-Wright, B. Kacar, C. Lineweaver, S. Rafkin, A. Soto, V. Thouzeau and members of the OCAV Project at PSL University and of NASA's Nexus for Exoplanet System Science (NExSS) research coordination network. We thank J. Kasting for his help adapting and running the VPL's photochemical and climatic models. B.S. is very grateful to Eleanor Lutz for her open-access codes of beautiful Martian maps (https://github.com/eleanorlutz/topography_atlas_of_space).

This work is supported by France Investissements d'Avenir programme (grant numbers ANR-10-LABX-54 MemoLife and ANR-10-IDEX-0001-02 PSL) through PSL IRIS OCAV and PSL–University of Arizona Mobility Program. RF acknowledges support from the US National Science Foundation, Dimensions of Biodiversity (DEB-1831493), Biology Integration Institute-Implementation (DBI-2022070), Growing Convergence in Research (OIA-2121155), and National Research Traineeship (DGE-2022055) programmes; and from the United States National Aeronautics and Space Administration, Interdisciplinary Consortium for Astrobiology Research program (award number 80NSSC21K059).

**Author contributions:**

Conceptualization: BS, BC, RF, SM
Methodology: BS, AA, RF, SM
Investigation: BS
Formal analysis: BS
Visualization: BS, RF
Software: BS, SM
Supervision: RF, SM
Writing – original draft –: BS
Writing – review & editing –: BS, BC, AA, RF, SM

**Competing interests:**

Authors declare that they have no competing interests.

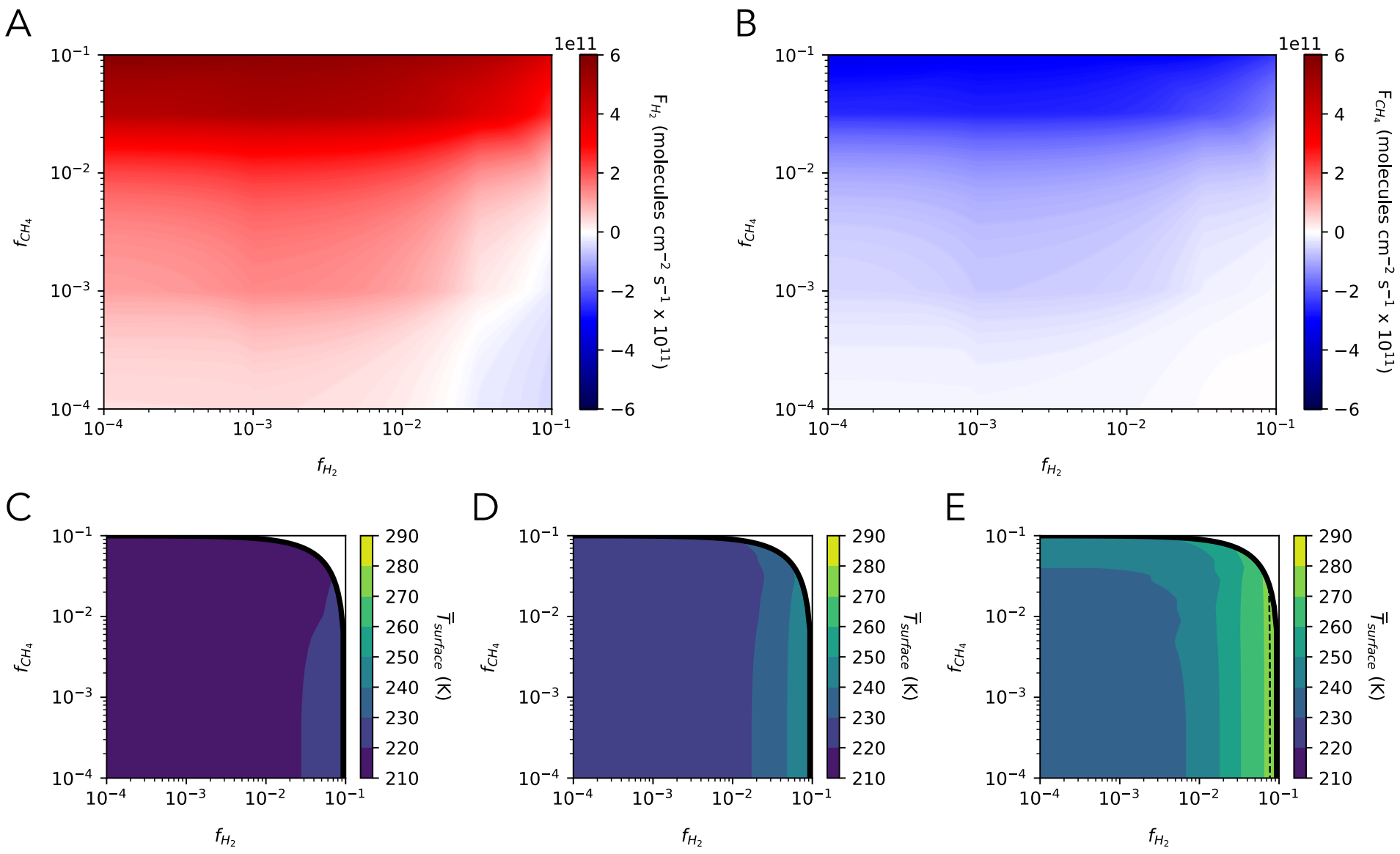

**Fig. 1 | Modeled photochemistry and climate of early Mars. A**, **B**, Rates (denoted by *F*) of destruction/production of $H_2$ and $CH_4$ as a function of $H_2$ and $CH_4$ mixing ratios (denoted by *f*). **C**, **D**, **E**, Average surface temperature ($\overline{T}_{surface}$) as a function of $H_2$ and $CH_4$ mixing ratios, $f_{H_2}$ and $f_{CH_4}$, for an atmospheric pressure of 0.5, 1, and 2 bars. Our simulations were constrained by $f_{H_2} + f_{CH_4} < 0.1$ (thick black boundary curve in **C**, **D**, **E**). Dashed black line in **E**: $\overline{T}_{surface}$ = 273 K. See Methods for details.

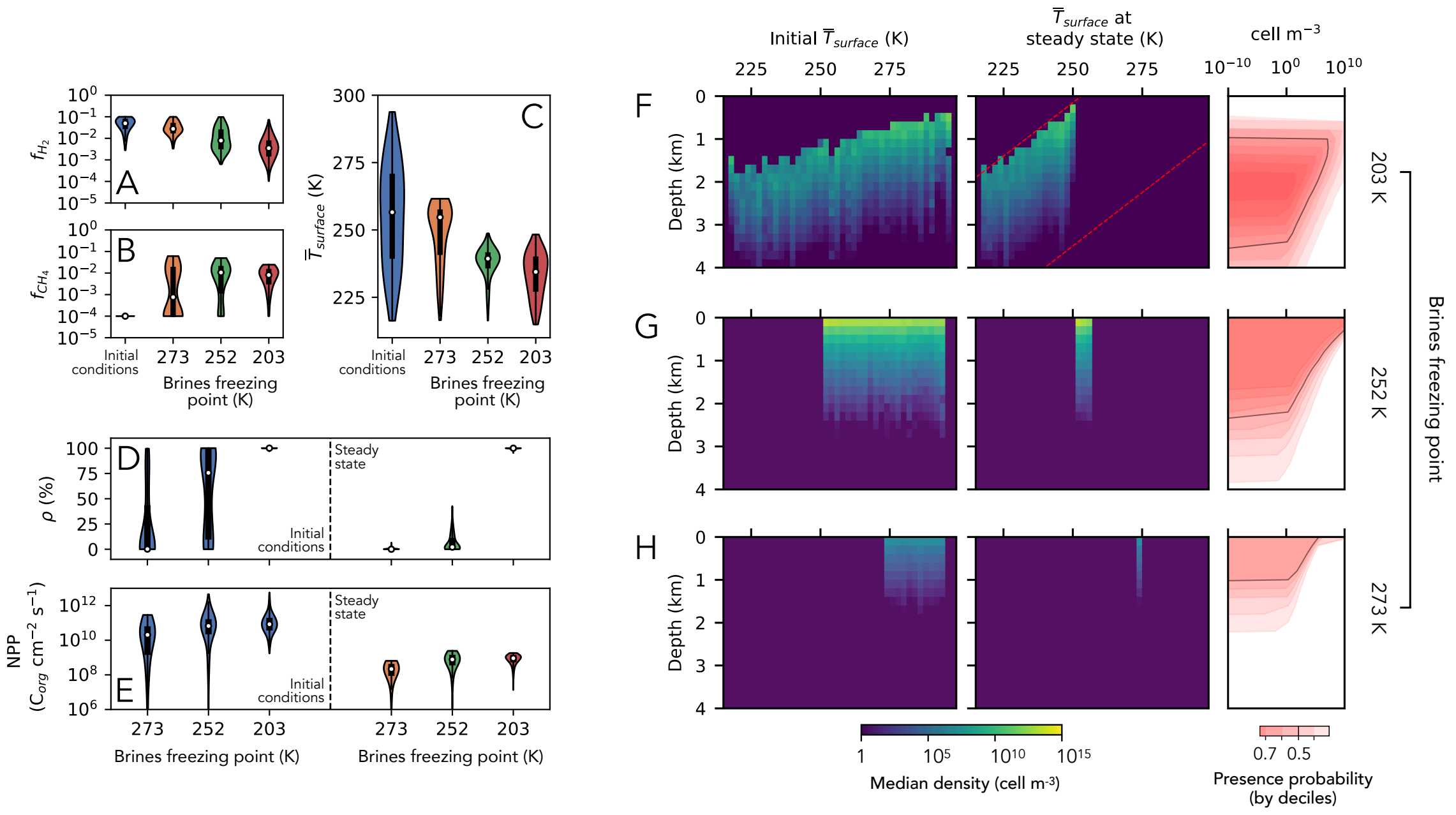

**Fig. 2 | Initial and steady state characteristics of Noachian Mars under the influence of hydrogenotrophic methanogens,** for brines freezing at 203, 252 and 273 K. **A, B,** Atmospheric composition. **C**, Average surface temperature. **D**, Ice coverage. **E**, Biomass production (in molecules of carbon fixed as biomass cm$^{-2}$ s$^{-1}$, averaged over the whole Martian surface, only when Mars is habitable). **F-H,** Depth profile of the subsurface methanogenic ecosystem at steady-state (median cell density in cells m$^{-3}$) as a function of the initial average surface temperature (*left*) and of the steady state average surface temperature (*middle*) and corresponding vertical probabilistic distribution of cell density (*right*). In **A-E**, the distributions corresponding to the initial characteristics of Mars are plotted in blue, and the distribution at steady state in orange, green, and red, when brines freeze at 273, 252, and 203 respectively. The white dots correspond to the median of the distrubtions, the thick black vertical lines correspond to their interquartile and the thin ones to their whole range. The red dotted lines in **F** correspond to the median depths at which temperature equals the lower and upper limits to viability, 253 and 320 K respectively. To resolve Mars' initial features, we followed previous work[14-16] and assumed atmospheric pressure ranging from 0.5 to 3 bars, a volume mixing ratio of $H_2$ , $f_{H_2}$, from 3,000 ppm to 0.1 corresponding to a volcanic outgassing rate of $10^{10}$ to 2 $10^{12}$ molecules cm$^{-2}$ s$^{-1}$, and a low $f_{CH_4}$ of 100 ppm corresponding to a production rate through serpentinization of 8 $10^8$ to $10^{10}$ molecules cm$^{-2}$ s$^{-1}$. The rest of the atmosphere was taken to be 95% $CO_2$ and 5% $N_2$[8]. We drew the characteristics of the crust (porosity, tortuosity and temperature profile) from the same ranges as in ref[9] and infer the posterior distributions of the depth profiles of temperature and diffusivity of the atmospheric gases (Extended Data Fig. 1). See Methods for more detail.

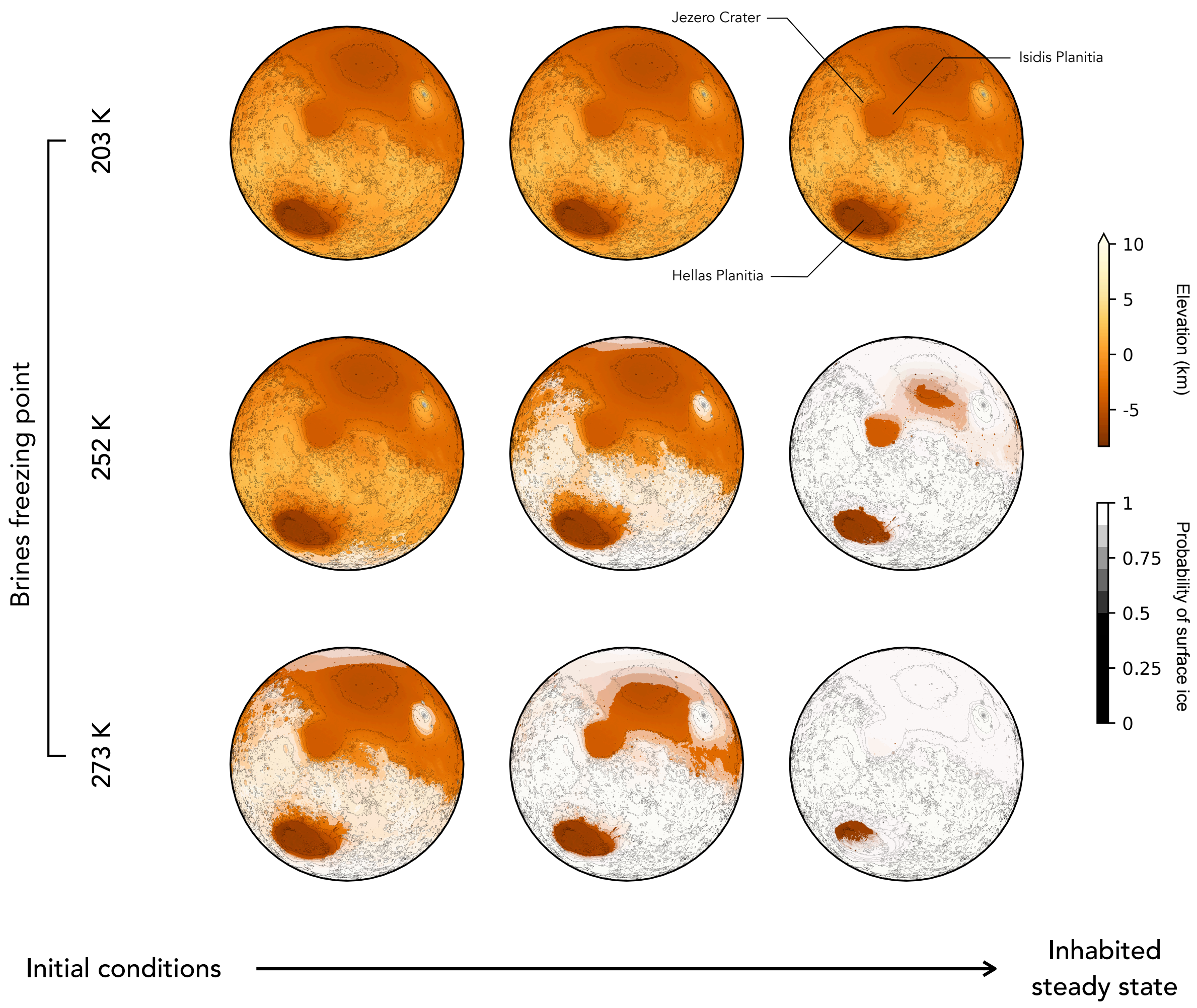

**Fig. 3 | Median evolution of the ice coverage of Noachian Mars under the influence of hydrogenotrophic methanogens** for brines freezing at 203 K (*top*), 252 K (*middle*), and 273 K (*bottom*). The orange color-scale represents elevation. The super-imposed white shaded areas correspond to the probability (from 0.5 to 1 by steps of 0.1) of surface ice. See Supplementary Video 1 for an animated version of the 252 K scenario.

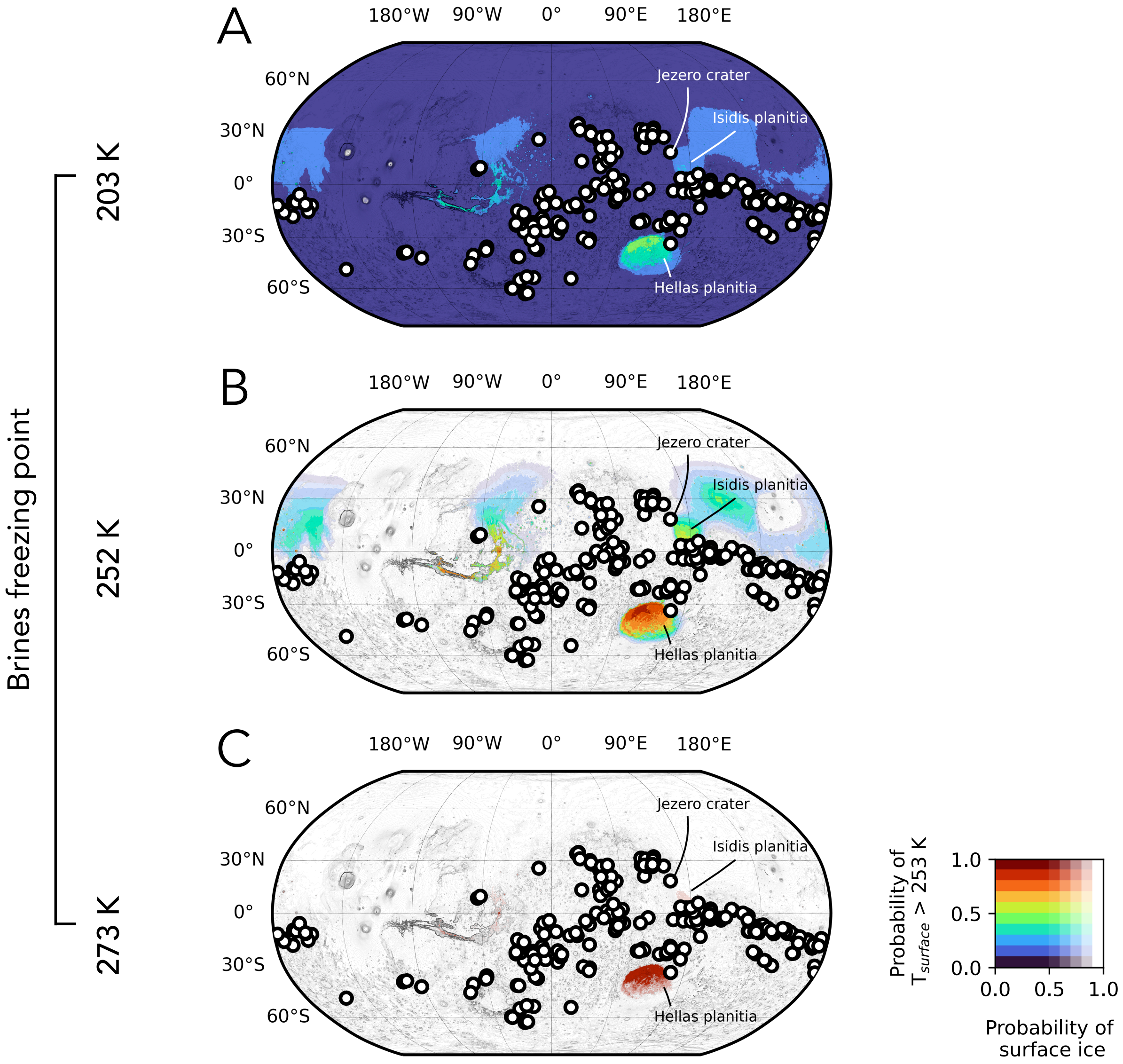

**Fig. 4 | Steady state distribution of habitable conditions and ice on the surface of Noachian Mars.** Spatial projection of the probability of surface ice (shades of white, from 0.5 to 1 by step of 0.1) superimposed on the probability of habitable surface temperature (i.e., $T_{surface}$>253 K, represented by the color gradient), for brines freezing at 203 K (**A**), 252 K (**B**), or 273 K (**C**). Open circles indicate the Noachian lakes distributed along the South-North dichotomy. See Methods for more detail.

# Supporting Information for "Early Mars' habitability and global cooling by $H_2$-based methanogens"

**Supplementary Table 1. Ranges of parameter values used in the atmosphere and crust models.**

| Planetary features | Symbol (units) | Value or Range | Distribution type |
|---|---|---|---|
| ***Initial atmosphere*** (ref[14-16]) | | | |
| Atmospheric pressure | p (bar) | [0.5 – 3] | uniform |
| $H_2$ mixing ratio | $f_{H_2}$ | [$3.10^{-3}$ – 0.1] | uniform |
| $CH_4$ mixing ratio | $f_{CH_4}$ | $10^{-4}$ | |
| $CO_2$ mixing ratio | $f_{CO_2}$ | [1 - ($f_{H_2}$+$f_{CH_4}$)] × 0.95 | |
| $N_2$ mixing ratio | $f_{N_2}$ | [1 - ($f_{H_2}$+$f_{CH_4}$)] × 0.05 | |
| ***Crust characteristics*** (ref[6,7]) | | | |
| Surface pore radius | $r(0)$ (cm) | [$10^{-4}$ – $10^{-3}$] | log-uniform |
| Depth of pore closure | $z_{max}$(km) | [5 – 20] | uniform |
| Surface porosity | $\epsilon(0)$ | [0.2 – 0.6] | uniform |
| Surface tortuosity | $\tau(0)$ | [1.5 – 2.5] | uniform |
| Temperature gradient | $a_T$ (K $km^{-1}$) | [10 – 40] | uniform |

**Supplementary Table 2. Biological parameters.**

| Parameter | Notation | Value or expression | Unit | References |
|---|---|---|---|---|
| Cell radius | $S_C$ | $10^{a_r+b_rT}$ | µm | ref[12] |
| | $a_r$ | -13.23 | dimensionless | ref[12] |
| | $b_r$ | 0.0431 | dimensionless | ref[12] |
| Cell volume | $V_C$ | $\frac{4}{3}\pi {S_C}^3$ | $\mu m^3$ | |
| Cellular carbon content | $Q_{C_{org}}$ | $18 \times 10^{-15}\ {V_C}^{0.94}$ | mol $C_{org}$ $cell^{-1}$ | ref[39] |
| Maximum metabolic rate | $q_{max}$ | $e^{a_q+b_qT}{V_C}^{c_q}$ | mol $cell^{-1}$ $d^{-1}$ | |
| | $a_q$ | - 55.76 | dimensionless | Ref[40] |
| | $b_q$ | 0.1 | dimensionless | ref[41] |
| | $c_q$ | 0.82 | dimensionless | ref[42,43] |
| Half-saturation constant | $K$ | $10^{-8}$ | mol $L^{-1}$ | ref[44,45] |
| Maintenance rate | $E_m$ | $e^{a_E+b_ET}{V_C}^{c_E}$ | kJ $d^{-1}$ | |
| | $a_E$ | - 43.54 | dimensionless | Ref[40] |
| | $b_E$ | 0.08 | dimensionless | ref[46] |
| | $c_E$ | 0.67 | dimensionless | ref[47] |
| Decay rate | $k_d$ | 0.5 | $d^{-1}$ | |
| Basal mortality rate | $m$ | 0.1 | $d^{-1}$ | |

**Supplementary references**

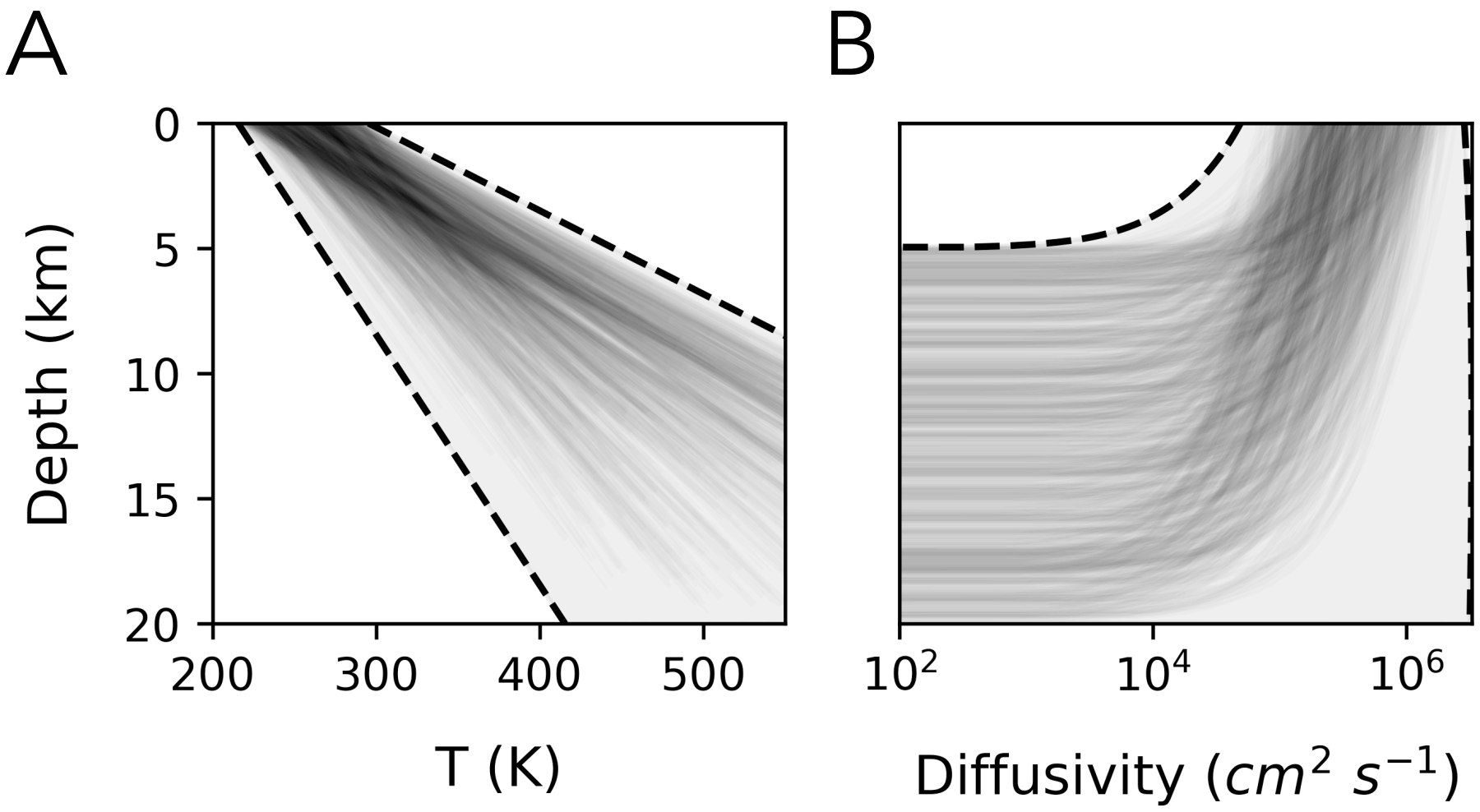

**Extended Data Fig. 1: Simulated depth profiles of (A) temperature and (B) diffusivity in Mars' Noachian regolith.** Gray areas bounded by dashed lines represent the entire space in which the depth profiles can exist. Each line (here 2,000 in total) represents one specific profile simulated for one set of parameters drawn from the ranges given in Supplementary Table 1.

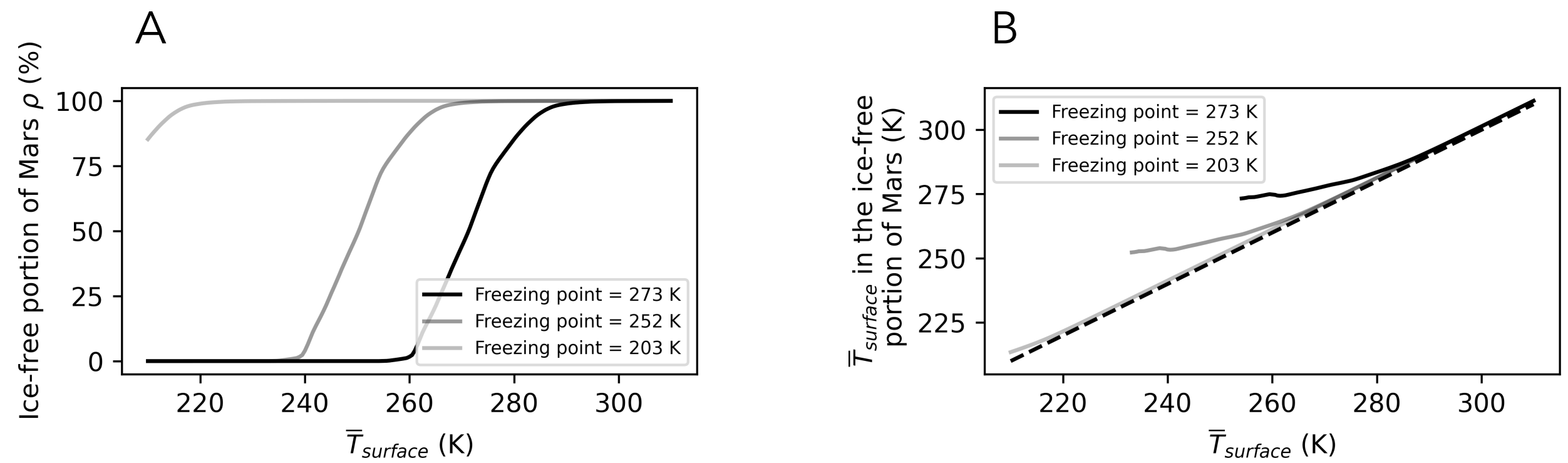

**Extended Data Fig. 2: Ice-free surface fraction, *ρ*, (A) and average temperature in the corresponding region (B).** Ice coverage and average surface temperature are evaluated across the spatial projection of Mars average temperature distribution (see Methods). The black dotted line in B is the first diagonal corresponding to the planetary averaged surface temperature $\overline{T}_{surface}$.

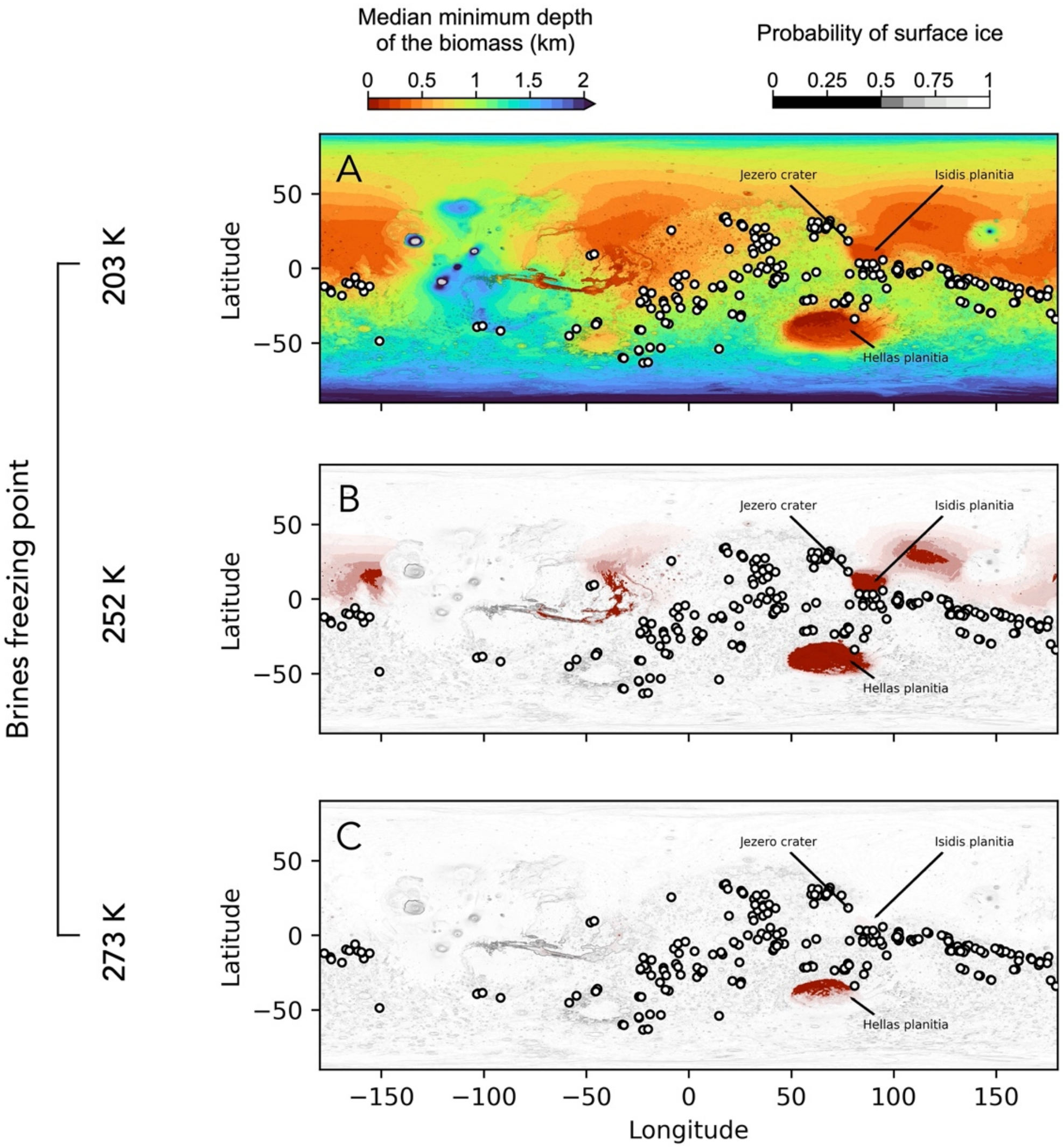

**Extended Data Fig. 3: Surface and vertical distribution of a putative hydrogenotrophic methanogenic biosphere on Noachian Mars.** Spatial projection of the median minimum depth of this biomass occurrence for three values of brines' freezing point of 203 K (A), 252 K (B), and 273 K (C). The white shaded areas correspond to the probability (from 50% to 90% by steps of 10%) of ice-coverage superimposed to the maps by transparency. Open circles indicate the Noachian lakes.